\documentclass{aa}
\usepackage{graphicx}
\usepackage{txfonts}
\begin{document}
   \title{Properties of dust in early-type galaxies}

   \subtitle{}
  
   \author{M. K. Patil
	\inst{1}$^\star$
	\and 
S. K. Pandey
	\inst{2}\thanks{Visiting Associate of IUCAA, Pune,
India}
	\and 
D. K. Sahu
	\inst{3} 
	\and 
Ajit Kembhavi
	\inst{4}
        }

   \offprints{S.K.Pandey}

   \institute{School   of   Physical  Sciences, S.R.T.M. University, Nanded, 431 606,  India\\
\email{patil@iucaa.ernet.in}
\and 
School of  Studies in  Physics, Pt. Ravishankar Shukla University, Raipur,  492 010, India\\
\email{skp@iucaa.ernet.in}
\and 
Indian Institute of  Astrophysics, Bangalore, 560 034,  India\\
\email{dks@crest.ernet.in}
\and 
Inter University Centre for Astronomy \& Astrophysics (IUCAA), Post  Bag   4,  Ganeshkhind,  Pune,   411  007,  India\\
\email{akk@iucaa.ernet.in}
    }

   \date{Received: 25 May 2005 / Accepted: 16 August 2006}

   \abstract{
We report optical extinction properties of dust for a sample of 26 early-type galaxies based on the analysis of their multicolour CCD observations. The  wavelength dependence of dust extinction for  these galaxies is determined and the extinction curves are found to run parallel to the  Galactic extinction curve, which implies that the properties of dust in the extragalactic environment are quite similar to those of the Milky Way. For the sample galaxies, value of the parameter $R_V$, the ratio of total extinction in $V$ band to selective extinction in $B$ \& $V$ bands, lies in the range 2.03 - 3.46  with an average of 3.02, compared to its canonical value of 3.1 for the Milky Way. A dependence of $R_V$ on  dust morphology of the host galaxy is also noticed in the sense that galaxies with a  well defined dust lane show  tendency to have smaller $R_V$ values compared to the galaxies with disturbed dust morphology. The dust content of these galaxies estimated using total optical extinction is found to lie in the range  $10^4$ to $10^6\,\rm M_{\sun}$, an order of magnitude smaller than those derived from IRAS flux densities, indicating that a significant fraction of dust intermixed with stars remains undetected by the  optical method. We  examine the relationship between dust mass derived from IRAS flux and the X-ray luminosity of the host galaxies.The issue of the origin of dust in early-type galaxies is also discussed.
   \keywords{Galaxies: elliptical and lenticular, cD --  Galaxies: ISM  -- ISM : dust, extinction }
   }
\titlerunning{Dust properties in early-type galaxies}
\authorrunning{M.K. Patil et al.}
   \maketitle
%
\section{Introduction} 
Recent imaging surveys using both ground and space-based telescopes across the electromagnetic spectrum have amply demonstrated that early-type galaxies, ellipticals (E) and lenticulars (S0) contain complex, multi-phase interstellar medium (ISM). In particular, a large fraction ($\sim$ 50 - 80\%) of E-S0 galaxies, at least in our local universe, are now known to possess dust features in a  variety of morphological forms, as revealed from high quality optical imaging, thanks to the availability of sensitive array detectors (Goudfrooij et al. 1994b,c; van Dokkum \& Franx 1995; Tran et al. 2001; and  references therein). Dust  attenuates the radiation at optical wavelengths and  re-radiates at infra-red (IR) wavelengths, especially in far-infrared (FIR) bands. FIR emission from  early-type galaxies detected using the {\it Infra Red Astronomical Satellite} (IRAS) (Knapp et al. 1989), and now with the {\it Infrared Space Observatory} (ISO) and Spitzer Space Telescope is attributed to thermal emission from dust having a wide variety of temperatures (Leeuw et al. 2004; Temi et al. 2004;  Xilouris et al. 2004). The dust masses derived from ISO data are at least  an order of  magnitude larger than those estimated using IRAS fluxes, which in turn have been found to be on average an order of magnitude higher than the dust mass estimated using optical extinction.

The study of physical properties of dust such as extinction, reddening and polarization help in deriving important information about the total dust content, size of the dust grains responsible for extinction and its variation with  environment, metallicity, star formation history and redshift of the host galaxy.  Dust properties in galaxies at high redshift are noticeably different from those found at later cosmic times, because the dust grains are smaller in size due to their different formation history (through type II Supernovae) and also due to the short time available for accreting heavy atoms and coagulation with other grains (Maiolino et al. 2004). Thus, detailed study of the dust properties in the extragalactic environment provides useful clues not only for understanding the origin and fate of dust in external galaxies,  but also for the subsequent evolution of host galaxies.

Our knowledge of  dust properties  largely relies on the interaction of dust particles with the electromagnetic radiation i.e., on attenuation and scattering of starlight, collectively known as {\em extinction}, and re-radiation by dust at longer wavelengths. The spectral dependence of extinction, termed as {\em extinction curve}, is found to depend strongly on the  composition, structure and size distribution of the dust grains. Therefore, the first step in understanding the physical properties of dust in the extragalactic environment is to derive the extinction curves for the sample galaxies. For our own Galaxy, the $\lambda$-dependence of extinction is derived by comparing the spectral energy distribution (SED) of a pair of stars of identical spectral and luminosity class with and without dust in front of them (Massa et al. 1983). Any difference in the measured magnitude and colour of these stars is attributed to the dust extinction. The standard extinction law thus derived  is found to be uniformly applicable in our Galaxy  from optical to near-infrared wavelengths. Over the  UV to far-IR wavelength range  the standard extinction law is characterized by a single parameter $R_V$, the ratio of total extinction $A_V$ in $V$ band to the selective extinction in $B\, \& \,V$ bands i.e., $E(B-V)$, with  $R_V =3.1$ (Savage \& Mathis 1979; Rieke \& Lebofsky 1985; Mathis 1990). However, within the Milky Way, the  value of $R_V$ is  found to vary between 2.1 to 5.6 depending on the line of sight (Valencic et al. 2004). Similarly, the extinction curves have also been derived for a few neighbouring spiral galaxies using this method and the  $R_V$ value is found to vary considerably for these galaxies (Brosch 1988), and even  within a galaxy from one location to another.

In the case of external galaxies, as individual stars cannot be resolved, the above  method is not applicable; instead, a variety of indirect methods have been proposed to establish the $\lambda$-dependence of the dust extinction, and to estimate $R_V$ values. In one of the widely used methods, comparison of light distribution of the original galaxy with its dust-free smooth model gives an estimate of the extinction caused by  dust present in the original galaxy. The extinction law for external galaxies can be deduced by performing this exercise at different wavelengths. The absence of spiral arms, \textsc{H ii} regions and other inhomogeneities results in a  fairly smooth light distribution in early-type galaxies. This  allows one to easily and  accurately construct the dust-free model of the galaxy required to study dust extinction properties. Despite this rather convenient situation, dust extinction properties have been investigated only for a handful of galaxies, most notably by Brosch \& Loinger (1991), Brosch \& Almoznino (1997), Goudfrooij et al. (1994b,c) and extended to a few more individual galaxies by Sahu et al. (1998), Dewangan et al. (1999), Falco et al. (1999), Keel \& White (2001), Motta et al. (2002) etc. In all these studies it has been observed that the optical extinction curves in the extragalactic environment closely resemble that of the Milky Way, with $R_V$ values comparable to the canonical value of 3.1. Dust being an important component of the ISM, it is essential to extend this kind of analysis to a larger sample of dusty early-type galaxies.

It has  been shown that the morphology of dust closely matches that of the ionized gas in a large fraction of galaxies ($\sim$ 50 - 80 \%) (Goudfrooij et al. 1994b, Ferrari et al. 1999), and in some cases with the X-ray emitting region too (Goudfrooij \& Trinchieri 1998), pointing to a possible physical connection between hot, warm and cold phases of ISM in early-type galaxies. In the multiphase ISM, dust grains can act as an efficient agent in transporting heat from hot gas to cold gas giving rise to the observed warm phase of ISM (Trinchieri et al. 1997). Again  a detailed investigation of dust properties is called for.

 We have an ongoing program of detailed surface photometric study of a large sample of early-type galaxies containing dust to investigate dust properties in the extragalactic environment and compare them with those of the Milky Way. This paper reports on dust properties in a sample of 26 early-type galaxies, based on their deep, broad band optical ($BVRI$) imaging observations. \S 2 describes the sample selection, observations and preliminary reduction of the acquired data, \S 3 describes the properties of dust for the  sample galaxies, in \S 4  results obtained are discussed. Our results are summarized in \S 5. Through out this paper we assume $\rm H_0$ = 50 km $\rm s^{-1}$ $\rm Mpc^{-1}$.


\section{Observations and data reduction}
 We have carried out multiband optical imaging observations of  26 {\em dusty} early-type galaxies (11 E and 15 S0) as a part of our ongoing program studying dust properties in a large sample of early-type galaxies taken from Ebneter \& Balick (1985), V\'eron-Cetty \& V\'eron (1988), Knapp  et al. (1989), van Dokkum  \& Franx (1995). The objects were chosen depending on the availability of observing nights and weather conditions, and as such no strict criterion has been applied for the  sample selection.  Table~\ref{obslog} gives  the global parameters such as coordinates, morphological type, heliocentric velocity, luminosity, size and environment of the target galaxies. 

\begin{table*}[!htb]
\caption[]{Global parameters of the program galaxies}
\centering
\label{obslog}
\begin{minipage}[t]{0.85\linewidth}
\scriptsize
 {\begin{tabular*}{1.0\linewidth}%
      {@{\extracolsep{\fill}}|cccccccl|}
      \hline\hline
      Object &RA & DEC &  Morph.& $B^{0}_{T}$ & $V_{helio}$ & size& CfA Group\\
      &  (J2000.0)& (J2000.0)& RC3(RSA)&  &$\rm{(km/s)}$ & (arcmin)& \\
      (1)&(2)& (3)& (4)&(5)& (6)& (7)& (8)\\
      \hline\hline
\object{NGC 524}&     01:24:47&   09:32:21& S0($S0/a$)& 11.17& 2509&  2.8x2.8&  GH13 (8), dominant in group\\
\object{NGC 984}&     02:34:43&   23:24:46& S0& 13.35& 4435&  3.0x2.0&  --\\
\object{NGC 1172}&    03:01:36&  -14:50:13& E1($S0_1$)& 12.56& 1502&  2.3x1.8 & --\\
\object{NGC 1439}&    03:44:50&  -21:55:21& E1($E1$)& 12.18& 1568&  2.5x2.3&  HG32 (21), Eridanus group\\
\object{NGC 2128}&    06:04:34&   57:37:40& S0& 12.66& 3226&  1.5x1.1&  --\\
\object{NGC 2534}&    08:12:54&   55:40:24& E1& 13.38& 3732&  1.4x1.2&  Isolated galaxy\\
\object{NGC 2563}&    08:20:35&   21:04:09& S0& 13.01& 4592&  2.1x1.5&  In dense part of Cancer cluster\\
\object{NGC 2672}&    08:49:22&   19:04:30& E1($E2$)& 12.12& 4164&  3.0x2.8&  Brightest in group, pair with N2673\\
\object{NGC 2693}&    08:56:59&   51:20:51& E3($E2$)& 12.61& 4923&  2.6x1.8&  GH39 (4)\\
\object{NGC 2907}&    09:31:36&  -16:44:09& S0/a($S0$)& 12.35& 1892& 1.8x1.1& Brightest in group, several dEs companion\\
\object{NGC 2911}&    09:33:46&   10:09:09& S0($S0$)& 12.25& 3062&  4.1x3.2&  GH47 (6)\\
\object{NGC 3078}&    09:58:24&  -26:55:34& E2($E3$)& 11.94& 2283&  2.5x2.1&  HG29(4), in cluster\\
\object{NGC 3489}&    11:00:18&   13:54:08& S0($S0/a$)& 11.15& 613 &  3.5x2.0&  GH68 (23)/HG56 (30), Leo grp\\
\object{NGC 3497}&    11:07:18&  -19:28:21& S0& 13.03& 3672&  2.6x1.4&  In cluster, interacting pair\\
\object{NGC 3585}&    11:13:17&  -26:45:20& E6($E/S0$)& 10.64& 1206&  4.7x2.6&  --\\
\object{NGC 3599}&    11:15:27&   18:06:46& S0& 12.70& 713 &  2.7x2.1&  GH77 (13)\\
\object{NGC 3665}&    11:24:43&   38:45:47& S0($S0$)& 11.69& 2072&  2.5x2.0&  GH79 (4), pair with N3658\\
\object{NGC 3923}&    11:51:02&  -28:48:23& E4($E/S0$)& 10.62& 1487&  5.9x3.9&  HG28 (4), at least 7 dEs\\
\object{NGC 4459}&    12:29:00&   13:58:46& S0($S0$)& 11.21& 1154&  3.5x2.7&  GH106 (248)\\
\object{NGC 5363}&    13:56:07&   05:15:19& I0($S0$)&  --  & 1113&  4.1x2.6&  HG55 (3), pair with N5364\\
\object{NGC 5485}&    14:07:11&   55:00:08& S0($S0$)& 12.31& 2107&  2.3x1.9&  HG77 (4), pair with N5484, N5486\\
\object{NGC 5525}&    14:15:39&   14:17:02& S0& 13.78& 5569&  1.4x0.9&  GH134 (8)\\
\object{NGC 5898}&    15:18:13&  -24:05:49& E0($S0$)& 11.92& 2153&  2.2x2.0&  In pair with NGC 5903,\\
& & & & & & &  several dEs, dS0s companion\\
\object{NGC 5903}&    15:18:36& -24:05:06& E2($E/S0$)&12.20& 2565&  2.7x2.1&  In pair with NGC 5898,\\
& & & & & & &  several dEs, dS0s companion\\
\object{NGC 7432}&    22:58:06&   13:08:00& E& 14.41& 7615&   1.5x1.2&  In pair\\
\object{NGC 7722}&    23:38:41&   15:57:15& S0&  13.19&  4197& 1.7x1.4& In pair with a spiral\\
\hline\hline
\end{tabular*}}
\begin{list}{}{}
\item[] \textsc{Notes}: Column (2) and (3) list galaxy co-ordinates, Column (4) lists morphological classification of the galaxies, RSA type is given in the parenthesis. The total, corrected blue luminosity of the program galaxies are listed in Column (5), while Column (6) lists heliocentric velocity  and Column (7) optical size of the galaxies, all taken from RC3 (de Vaucoulers et al. 1991). Column (8) lists CfA group membership HG or GH listed by Huchra \& Geller (1982; HG82) and Geller \& Huchra (1983; GH83), number in the parenthesis represents number of galaxies in a given group.
\end{list}
\end{minipage}
\end{table*}
Deep CCD images  of the program  galaxies  were obtained  using  various observing  facilities available in India,  during  December 1998 to August 2003. Table~\ref{instru} gives details of the instruments used during different observing runs and  Table~\ref{obslog1} gives the log of   observations. Except for the IAO observing run, where Bessel $U, B, V, R, I$ filters were used, observations were made in Johnson $B, V$ and Cousins $R, I$ filters.  Generally, exposure times were adjusted so as to achieve roughly equal signal-to-noise (S/N) ratio for a galaxy in different bands. Apart from the object   frames, several calibration  frames such as  bias, twilight sky flats  etc. were also taken in each observing run. For photometric calibration of the data, open cluster M67 and standard stars from Landolt's list  (Landolt 1992) were observed during photometric nights.

\begin{table*}[htb]
\caption[]{Details of telescopes and instrumentation}
\centering
\label{instru}
\scriptsize
\begin{minipage}[t]{0.7\linewidth}
{\begin{tabular*}{1.0\linewidth}%
{@{\extracolsep{\fill}}|lll lll ll|}\hline
Observing Run&           Dec.98&   Dec.99&   Mar.2k&  Mar.01&  Apr.01&  May 03&   Aug.03\\
\hline
Observatory&   ARIES&      ARIES&      VBO&       VBO&      VBO&     IAO&   IAO\\
Telescope size(in m)& 1.04&	 1.04&     2.34&      2.34&    2.34&    2.0&      2.0\\
CCD Type&      TEK&	 TEK&	   Phot&     Phot&    Phot&    Phot&     Phot\\
Format (\# pixels)& $1k\times 1k$& $2k\times 2k$& $1k\times 1k$& $1k\times 1k$& $1k\times 1k$& $1k\times 1k$& $1k\times 1k$\\
Binning&   $2\times2$& $2\times2$&  $1\times1$& $1\times1$& $1\times1$& $1\times1$& $1\times1$\\
Scale ($\arcsec$/bin)& 0.$\arcsec$73& 0.$\arcsec$73& $0.$\arcsec$72$& $0.$\arcsec$72$& $0.$\arcsec$72$& $0.$\arcsec$285$& $0.$\arcsec$285$\\
Field of view&  $13\arcmin\times13\arcmin$&  $13\arcmin\times13\arcmin$& $10\arcmin\times10\arcmin$& $10\arcmin\times10\arcmin$& $10\arcmin\times10\arcmin$& $5\arcmin\times5\arcmin$& $5\arcmin\times5\arcmin$\\
\hline
\end{tabular*}}
\begin{list}{}{}
\item[]\textsc{Notes}: Telescope - 1.04 $\doteq$ 1.04m Sampurnanand Telescope of Aryabhatta Research Institute of Observational Sciences (ARIES), Naini Tal;  $2.34\doteq 2.34m$ Vainu Bappu Telescope of V.B. Observatory (VBO), Kavalur; $2.0\doteq 2.0m$ Himalayan Chandra Telescope of Indian Astronomical Observatory (IAO), Mt. Saraswati, Hanle. \\
CCD type - TEK\,: Tektronics, Phot\,: Photometrics.
\end{list}
\end{minipage}
\end{table*}
\begin{table*}[htb]
\caption[]{Observing log}
\label{obslog1}
\centering
\begin{minipage}[t]{0.8\linewidth}
\scriptsize
 {\begin{tabular*}{1.0\linewidth}%
{@{\extracolsep{\fill}}|l|llll| cccc|c|}\hline\hline
Galaxy &   \multicolumn{4}{c|}{Exposure time (sec)}&  \multicolumn{4}{c|}{Seeing ($\arcsec$)}&  Notes\\
 &                      B& V& R& I& B& V& R& I&  \\
(1)& (2)& (3)& (4)& (5)& (6)& (7)& (8)& (9)& (10)\\
\hline
NGC 524  &  900(4) & 900(3) & 400(3) & 400(3) & 2.65 & 2.41 & 2.14 & 2.11 & \#1 \\
NGC 984  &  900(4) & 600(4) & 400(3)+200(1) & 300(4) & 3.01 & 2.72 & 2.71 & 2.91 & \#2 \\
NGC 1172 &  900(4) & 900(3) & 400(3) & 400(3) & 2.66 & 2.55 & 2.46 & 2.61 & \#2 \\
NGC 1439 &  600(4) & 600(3) & 200(4)  & 200(3)  & 2.55 & 1.93 & 1.86 & 1.95 & \#2 \\
NGC 2128 &  900(4) & 900(3) & 400(2)  & 400(2) & 2.81 & 2.46 & 2.39 & 2.09 & \#1 \\
NGC 2534 &  900(4) & 900(3) & 600(3) & 600(3) & 2.51 & 2.25 & 1.79 & 2.06 & \#1 \\
NGC 2563 &   -  & 600(4) & 400(3) & 400(3) &   -  & 4.98 & 4.52 & 4.32 & \#3 \\
NGC 2672 &  900(4) & 600(3) & 300(3)+200(1) & 300(3)  & 1.90 & 1.87 & 2.08 & 1.88 & \#2 \\
NGC 2693 &  900(4) & 600(3) & 360(4) &  300(4)& 3.30 & 3.33 & 3.07 & 2.96 & \#4 \\
NGC 2907 &  900(4) & 600(3) & 300(4) & 300(3)  & 3.46 & 2.61 & 2.57 & 2.46 & \#2 \\
NGC 2911 &  1200(3)+600(1) & 600(4) & 300(5) & 300(4) & 2.62 & 2.30 & 2.00 & 1.86 & \#1 \\
NGC 3078 &  600(4) &  -   & 300(2)  &   -  & 2.87 &  -   & 3.29 &  -   & \#4 \\
NGC 3489 &  500(3) & 400(3) & 400(1)  & 360(1)  & 2.97 & 2.75 & 2.77 & 2.87 & \#2    \\
NGC 3497 &    -  & 700(3) &  -   & 420(2)  &  -   & 2.38 &  -   & 2.41 & \#7 \\
NGC 3585 &  600(4) & 600(3) & 300(2)  &  -   & 3.16 & 3.07 & 2.90 &  -   & \#3 \\
NGC 3599 &  600(3) & 300(3)  & 300(3)  & 200(3)  & 2.59 & 2.54 & 2.47 & 2.21 & \#2    \\
NGC 3665 &  600(3) & 400(3) & 300(2)  &  -   & 2.11 & 1.96 & 1.82 &  -   & \#6 \\
NGC 3923 &  900(4) & 600(4) & 300(4) &  -   & 3.24 & 3.60 & 3.19 &  -   & \#3 \\
NGC 4459 &  900(4) & 600(3) & 360(4) &  -   & 2.50 & 2.41 & 2.20 &  -   & \#5 \\
NGC 5363 &  600(4) & 600(3) & 300(4) & 300(2)  & 2.12 & 1.96 & 2.42 & 1.82 & \#6 \\
NGC 5485 &  600(6) & 600(3) & 300(4) & 300(6) & 2.11 & 1.96 & 1.54 & 1.52 & \#7 \\
NGC 5525 &  600(6) & 500(4)+280(1) &   -  &  -   & 2.01 & 2.11 &   -  &  -   & \#7 \\
NGC 5898 & 800(3) & 400(3) & 300(2)  & 300(2)  & 3.25 & 2.61 & 3.06 & 2.89 & \#3 \\
NGC 5903 & 800(3) & 400(3) & 300(2)  & 300(2)  & 3.25 & 2.61 & 3.06 & 2.89 & \#3 \\
NGC 7432 & 600(6) & 600(3) & 300(6) & 300(6) & 1.41 & 1.37 & 1.25 & 1.48 & \#6 \\
NGC 7722 & 600(5) & 600(3) & 400(3) & 400(3) & 1.45 & 1.75 & 1.30 & 1.61 & \#4 \\
\hline\hline
\end{tabular*}}
\scriptsize
\begin{list}{}{}
\item[]\textsc{Notes to Table ~\ref{obslog1}:} Column (1) lists sample galaxies. Columns (2)-(5) list  exposure time in seconds for  B, V, R, \& I bands, respectively,  number in  parenthesis represents the number of exposures acquired in a given band. Seeing (in arcsec) in different bands are listed in columns (6)-(9). Column (10) lists  observing run and  instruments used; $\#1\doteq$ ARIES, 1998; $\#2\doteq$ ARIES, 1999; $\#3\doteq$ VBO, March 2000; $\#4\doteq$ VBO, March 2001; $\#5\doteq$ VBO, April 2001; $\#6\doteq$ IAO, May 2003; $\#7\doteq$ IAO, August 2003 (see Table ~\ref{instru}).
\end{list}
\end{minipage}
\end{table*}
Standard preprocessing steps such as bias subtraction and flat fielding were done using the standard tasks  available within IRAF\footnote{IRAF  is distributed  by the National  Optical Astronomy  Observatories (NOAO), which  is operated  by the  Association of  Universities,  Inc. (AURA) under  co-operative agreement with  the National  Science Foundation.}. Multiple  frames  taken in each filter were geometrically aligned to an accuracy  better than one tenth of a pixel by measuring centroids of several common stars in the galaxy frames, and were then combined  to  improve  the S/N ratio. This also enabled easy removal of cosmic ray events. The cosmic ray hits that were left after combining the frames  were further eliminated using the {\it cosmicrays} task in IRAF.  Sky background in the galaxy frame was estimated using  the box method (see, e.g., Sahu et al. 1998);   median of a 5$\times$5 pixels box at various locations in the frame, generally  away from the galaxy and not affected by the stars,  was estimated and  its mean value was taken as a measure of the  sky background which was then subtracted from the respective galaxy frame in the corresponding band. As the total field coverage of the CCDs used was large compared to the optical  size of the program galaxies, the box  method for sky estimation was found  suitable for the present study. Cleaned, sky subtracted $B, V, R, I$ images of individual galaxies  were convolved with a Gaussian function to match the seeing of the best frame  with that of  the  worst frame  to construct colour-index images of the galaxies.

\section{Properties of dust: the method and results}
\begin{figure*}[!htb]
\centering
\resizebox{0.65\textwidth}{!}{\includegraphics{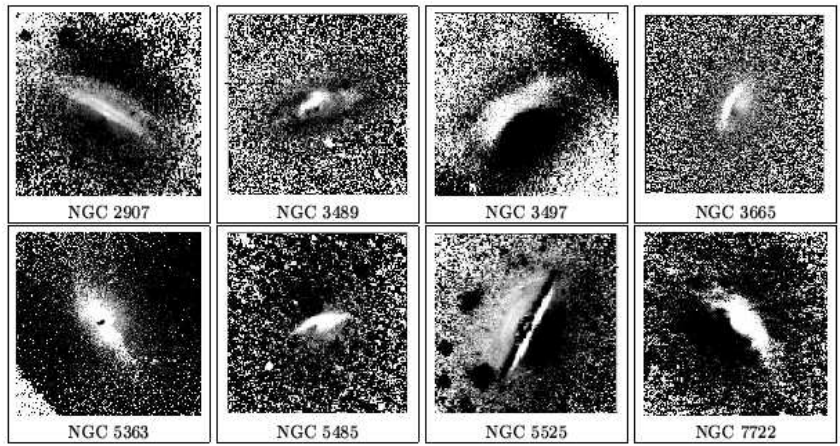}}
\caption{($B-V$) colour index maps of some of the prominent dust lane galaxies; brighter shade represents  dust occupied regions}
\label{fig:colormap}
\end{figure*}

Even though the program galaxies were known to contain dust features, we re-examined them using a variety of image processing techniques like quotient image, unsharp masking, colour-index image, etc., to confirm the presence of dust features and their spatial distribution. ($B-V$) and ($B-R$) colour index images  were constructed using geometrically aligned, seeing matched direct images. ($B-V$) colour index images of some of the dust lane galaxies are shown in  Fig.~\ref{fig:colormap}, where brighter shades represent  dust occupied redder regions.  Our analysis confirms the presence of dust in all the 26 galaxies studied here. Dust is present in variety of forms; five galaxies have  multiple dust lanes parallel to the major axis, twelve objects show a well-defined dust lane aligned either along major or minor axes, three galaxies show dust rings or arcs, while others show nuclear dust patches. NGC 2907 and NGC 7722 have at least four extended dust lanes running parallel to the optical major axis. NGC 5363 has two lanes, the  inner one is  short and  aligned along minor axis, while the  other one is extended and lies parallel to the major axis of the galaxy. 

\subsection{Extinction maps}
To estimate the  effect of dust extinction, we compare the light distribution in the original galaxy image with its dust free model. As early-type galaxies have a fairly smooth and  symmetric stellar light distribution with respect to the nucleus, one can easily construct  its smooth, dust free  model by fitting ellipses to the isophotes of observed image. This method has been used by a number of researchers (Brosch \& Loinger 1991; Goudfrooij et al. 1994c; Sahu et al. 1998, and references therein) to study extinction properties of dust in the extragalactic environment. 

We have fitted ellipses to the isophotes of the observed galaxy images using the {\em ellipse} fitting routine available in the STSDAS\footnote{STSDAS is distributed by the Space Telescope Science Institute, operated by AURA, Inc., under NASA contract NAS 5-26555.} package, which is based on a procedure described by Jedrzejewski (1987). 
\begin{figure*}[!htb]
\centering
\resizebox{0.72\textwidth}{!}{\includegraphics{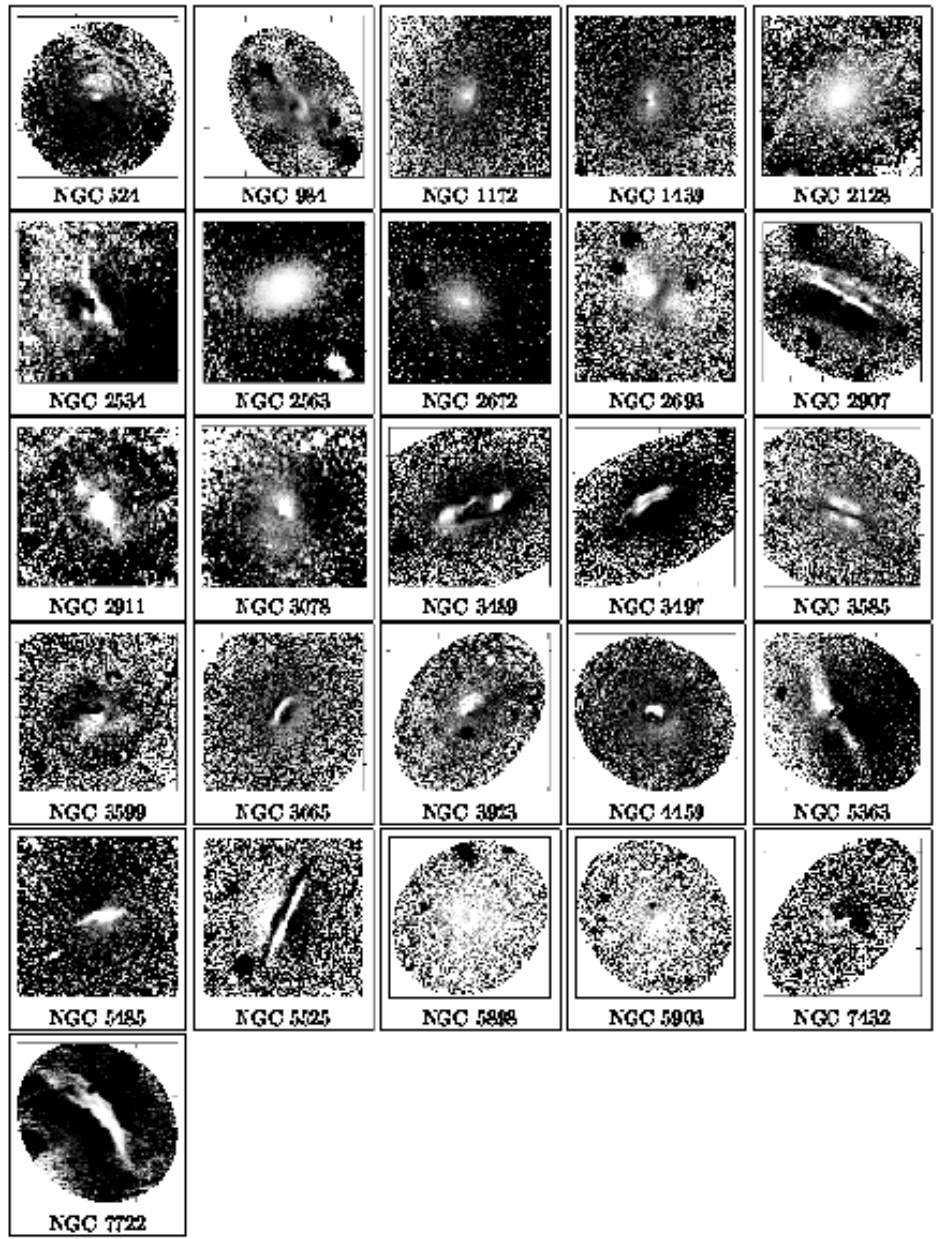}}
\caption{Extinction maps; brighter shades represent dust occupied regions. North is up and East is to right.}
\label{fig:extinmap}
\end{figure*}

Starting with  trial values of ellipticity, position angle, and centre coordinates of the galaxy image, an ellipse was fitted to the isophote at a given semi-major axis length  after masking all the obvious regions occupied by foreground stars and interacting galaxies, which were ignored during the ellipse fit. The fitting was continued by incrementing the semi-major axis length by 10\% until signal reaches 3$\sigma$ level of the background, and continued inward until the centre of the galaxy. A model image  constructed using the best fit ellipses was then subtracted from the original galaxy image and its residual image was generated. The residual image was examined and  regions occupied by dust and other hidden features were flagged and rejected in the next run of ellipse fitting. As the majority of galaxies from this sample contain complex dust lanes or patches passing through their centres, accurate estimation of centre coordinates was not possible. To minimize the error in this estimate, isophote fitting was first carried out  for $R$ or $I$ band images, which are least affected by dust extinction among the available bands and centre coordinates were determined by averaging those of the best fitted ellipses. Centre coordinates thus determined were kept fixed in the second run and the above described ellipse-fitting procedure was repeated. The same centre coordinates were used to fit ellipses to the isophotes in other bands. 

The ``dust free'' model of the program galaxies thus generated were used to construct ``extinction maps'' in magnitude scale using the relation
\begin{equation}
\rm A_\lambda=-2.5\times log \left[\frac{I_{\lambda,obs}}{I_{\lambda,model}}\right]
\end{equation}
where $A_{\lambda}$ represents the amount of extinction in a particular band $\lambda$ (= $B, V, R, I$) while $\rm I_{\lambda,obs}$  and $\rm I_{\lambda,model}$ represent ADU count levels in the original and model galaxies, respectively. 

Extinction maps for the sample galaxies  are shown in Fig.~\ref{fig:extinmap}, where brighter features represent  regions of higher optical depth associated with the dust extinction. Asymmetries seen in the dust features of NGC 5903 and NGC 7432 are due to the masking of foreground stars present near the centre of galaxies. 

\subsection{Extinction curves}
The next step involves the quantification of total extinction in each band and deriving the extinction curve. For this purpose, masks were set on the regions  occupied by dust in extinction maps generated above, and numerical values of $A_\lambda$ (with $\lambda=B,V,R~ \& ~I$) were extracted as the mean extinction within a square box of $5\times 5$ pixels (comparable to the size of the seeing disk) and the box was slided over the dust-occupied region in each galaxy. To avoid  seeing effects, we excluded nuclear regions (radius $\le 5\arcsec$) of the program galaxies from analysis. The  values of extinctions $A_\lambda (x,y)$ thus measured at different locations in individual bands were used to estimate the local values of selective extinction or colour excess $E(\lambda-V)$ = $A_ \lambda- A_V$ as a function of position across the dust-occupied region.

\begin{table*}[!htb]
\caption{$R_\lambda$ values and relative grain sizes}
\label{Rvtab}
\centering
\scriptsize
\begin{minipage}{0.8\linewidth}
\begin{tabular*}{1.0\linewidth}%
{@{\extracolsep{\fill}}|l|ccccc|}\hline\hline
Object&		$R_B$&	$R_V$&	$R_R$&	$R_I$&		$\frac{<a>}{a_{Gal}}$\\
(1)&   (2)&  (3)&  (4)& (5)&  (6)\\
\hline
NGC 524&	$4.46\pm 0.39$& $3.46\pm 0.16$&	$2.21\pm 0.47$&	$1.57\pm 0.45$&	$1.06\pm 0.02$\\
NGC 984&	$4.19\pm 0.36$& $3.19\pm 0.21$&	$2.43\pm 0.19$&	$2.06\pm 0.17$&	$1.08\pm 0.08$\\
NGC 1172&	$4.25\pm 0.17$& $3.25\pm 0.19$&	$2.39\pm 0.19$&	$1.95\pm 0.14$&	$1.07\pm 0.05$\\
NGC 1439&	$3.45\pm 0.21$& $2.45\pm 0.17$&	$2.01\pm 0.15$&	$1.76\pm 0.13$&	$0.93\pm 0.08$\\
NGC 2128&	$4.37\pm 0.28$& $3.37\pm 0.27$&	$2.06\pm 0.29$&	$1.59\pm 0.30$&	$1.04\pm 0.06$\\
NGC 2534&	$3.03\pm 0.19$& $2.03\pm 0.28$&	$1.61\pm 0.30$&	$1.18\pm 0.38$&	$0.80\pm 0.04$\\
NGC 2563&	$--$& 		$3.10\pm 0.00$&	$2.29\pm 0.74$&	$1.19\pm 0.51$&	$0.96\pm 0.03$\\
NGC 2672&	$4.27\pm 0.14$& $3.27\pm 0.28$&	$2.48\pm 0.14$&	$1.77\pm 0.15$&	$1.06\pm 0.02$\\
NGC 2693&	$4.46\pm 0.23$& $3.46\pm 0.26$&	$2.87\pm 0.45$&	$2.59\pm 0.36$&	$1.30\pm 0.04$\\
NGC 2907&	$3.25\pm 0.40$& $2.25\pm 0.48$&	$1.19\pm 0.29$&	$0.49\pm 0.23$&	$0.77\pm 0.03$\\
NGC 2911&	$3.84\pm 0.35$& $2.84\pm 0.29$&	$1.98\pm 0.31$&	$1.13\pm 0.37$&	$0.92\pm 0.04$\\
NGC 3078&	$4.10\pm 0.17$& $--$&		$2.39\pm 0.19$&	$--$&		$1.02\pm 0.03$\\
NGC 3489&	$4.38\pm 0.25$& $3.38\pm 0.21$&	$2.64\pm 0.19$&	$1.70\pm 0.15$&	$1.09\pm 0.02$\\
NGC 3497&	$--$& 		$3.10\pm 0.00$&	$--$&		$1.53\pm 0.65$&	$1.02\pm 0.02$\\
NGC 3585&	$4.45\pm 0.44$& $3.45\pm 0.33$&	$3.00\pm 0.24$&	$--$&		$1.17\pm 0.04$\\
NGC 3599&	$3.49\pm 0.15$& $2.49\pm 0.09$&	$2.12\pm 0.06$&	$1.57\pm 0.14$&	$0.93\pm 0.06$\\
NGC 3665&	$4.29\pm 0.20$& $3.29\pm 0.13$&	$2.33\pm 0.23$&	$--$&		$1.03\pm 0.02$\\
NGC 3923&	$4.16\pm 0.43$& $3.16\pm 0.39$&	$2.03\pm 0.16$&	$--$&		$0.98\pm 0.04$\\
NGC 4459&	$3.94\pm 0.24$& $2.94\pm 0.18$&	$2.26\pm 0.16$&	$--$&		$0.90\pm 0.02$\\
NGC 5363&	$3.78\pm 0.27$& $2.78\pm 0.16$&	$2.18\pm 0.36$&	$1.89\pm 0.28$&	$0.92\pm 0.04$\\
NGC 5525&	$4.15\pm 0.29$& $3.15\pm 0.17$&	$--$&	$--$&		$0.99\pm 0.01$\\
NGC 5898&	$4.15\pm 0.30$& $3.15\pm 0.46$&	$2.46\pm 0.29$&	$1.84\pm 0.41$&	$1.04\pm 0.03$\\
NGC 5903&	$4.07\pm 0.30$& $3.07\pm 0.22$&	$2.28\pm 0.39$&	$2.01\pm 0.34$&	$1.05\pm 0.08$\\
NGC 7432&	$3.89\pm 0.28$& $2.89\pm 0.21$&	$2.46\pm 0.19$&	$1.94\pm 0.22$&	$1.05\pm 0.09$\\
NGC 7722&	$3.89\pm 0.25$& $2.89\pm 0.20$&	$2.03\pm 0.14$&	$1.78\pm 0.10$&	$0.94\pm 0.02$\\
\hline
&	$R_U$~~~~~~~~~~~~~~~$R_B$& $R_V$&	$R_R$& $R_I$&	 \\
NGC 5485& $4.38\pm 0.24$~~~~~$3.65\pm 0.21$& $2.65\pm 0.17$& $1.86\pm 0.12$& $1.40\pm 0.18$&  $0.92\pm 0.04$\\
\hline
{\bf The Galaxy}&     4.10&          3.10&  2.27&  1.86& 1.00\\
\hline\hline
\end{tabular*}
\begin{list}{}{}
\item[]\textsc{Notes on Table~\ref{Rvtab}}: Column (2) to (5) lists the $R_\lambda\ \left[\equiv \frac{ A_\lambda}{E(B-V)}\right]$ values with $\lambda= B,\, V,\, R,\, \& \, I$. Column (6) lists the relative (mean) grain size in the program galaxies with respect to that responsible for the Galactic curve. NGC 5485 represents data in five bands (U,B,V,R \& I). For comparison, Galactic $R_\lambda$ values are listed in the last row.
\end{list}
\end{minipage}
\end{table*}

A linear regression  fit was performed between  various local values of total extinction ($A_B, A_V, A_R, A_I$) and the slopes of the best fits were assigned to be the average slope of $A_x$ versus $A_y$ (where $x,y = B, V, R\,\&\, I$; $x\neq y$) and the reciprocal slope of $A_y$ versus $A_x$ (Goudfrooij et al. 1994b; Sahu et al. 1998; Dewangan et al. 1999). Likewise, slopes of the fitted lines of the regression for $A_\lambda$ ($\lambda=B,V, R\, \& \, I$) and $E(B-V)$ were also derived. The best fitting slopes were used to derive $R_\lambda\ \left[\equiv \frac{ A_\lambda}{E(B-V)}\right]$ for the dust occupied regions in the program galaxies and are listed in Table~\ref{Rvtab} along with their associated errors.  $R_\lambda$ values for the Milky Way taken from Rieke \& Lebofsky (1985) are also listed in the table for comparison. 

\begin{figure*}[!htb]
\centering 
\resizebox{0.7\textwidth}{!}{\includegraphics{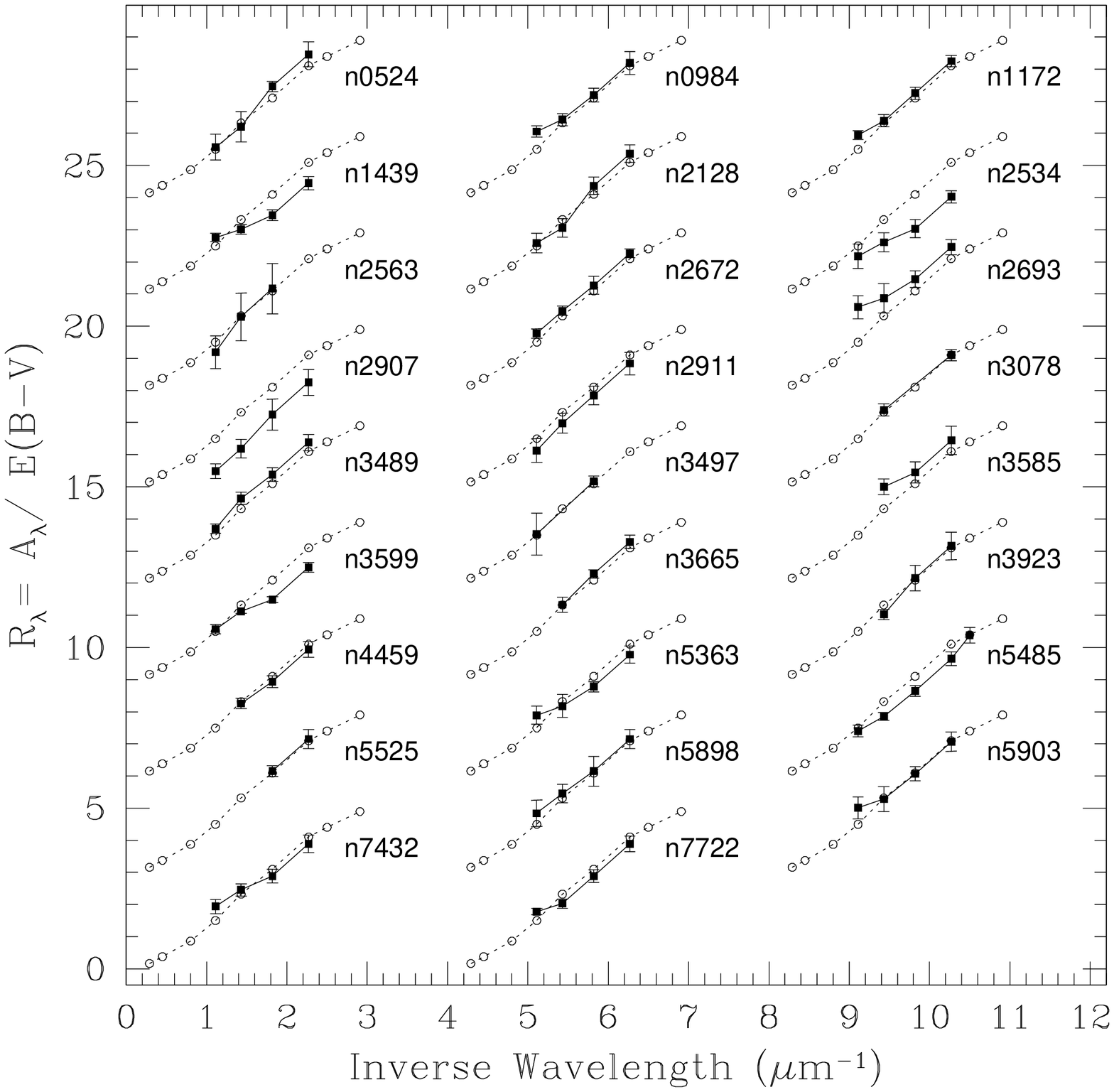}}
\caption{Extinction curves for the program galaxies (filled squares, solid lines) along with the canonical curve for the Galaxy (open circles, dashed lines) for comparison. The offset (shift of the curves along the X or Y axis in order to separate them) for the X-axis is 4, that for the Y-axis is 3.}
\label{extincurves}
\end{figure*}

The {\em extinction curves} for the program galaxies are given in Fig.~\ref{extincurves} along with that for the Milky Way. These figures demonstrate that the extinction curves for the majority of galaxies run parallel to that of the Galactic curve, except for a few cases (NGC 1439, 2534, 2693, 3585, 5363) for which we see ``concave'' extinction curves. The fact that, on average, extinction curves for the sample galaxies  are parallel to that of our Galaxy implies that the dust extinction properties in the extragalactic environment are similar to those of the Milky Way. However, $R_V$ for program  galaxies is found to be different from the canonical value of 3.1 for the Milky Way, as evident from Table~\ref{Rvtab}. 

Assuming that the chemical composition of  dust grains in the extragalactic environment is similar to that of our Galaxy, smaller (larger) $R_V$ values imply that the dust grains responsible for interstellar reddening are smaller (larger) than the grains in our Galaxy. This allows one to compute the relative grain size in the sample galaxies by making use of available models of chemical composition and shape of the dust grains. Out of several ``dust models'' a  two-component model consisting of spherical graphite and silicate grains, with an  adequate mixture of sizes  (Mathis et al. 1977)  is able to explain the observed  extinction curves in the Milky Way as well as in the Local Group (e.g., Clayton et al. 2003).  This model assumes uncoated refractory particles having a power law size distribution  $n(a)=n_0\,\,a^{-3.5}$, where $a$ represents grain radius. As the present study is restricted to the optical regime, the model proposed by Mathis et al. (1977) serves as a good approximation for this purpose.

Fig.~\ref{extincurves} shows that $R_\lambda\ \left[\equiv \frac{ A_\lambda}{E(B-V)}\right]$ varies linearly with the inverse of the wavelength and is consistent with the result $Q_{ext} \propto \lambda^{-1}$ expected for small grains i.e., for $x\le 1$, where $x = \frac{2 \pi a }{\lambda}$; $a$ is the grain radius and $Q_{ext}$ is the extinction efficiency factor. Thus, for a given $Q_{ext}$ value, the mis-match between the extinction curve for the Milky Way  and those for the program  galaxies is attributed to the difference  in  grain size between the program galaxies and the Milky Way. Therefore, one can estimate the relative grain size in such cases by shifting the observed extinction curve along ${\lambda}^{-1}$ axis until it best matches the Galactic extinction curve (cf. Goudfrooij et al. 1994c), in the sense that the extinction curve lying below (above) the Galactic curve will correspond to  smaller (larger) grain size relative to that of Galactic dust grains. The relative grain sizes thus derived for the sample galaxies are listed in Column 6 of Table~\ref{Rvtab}.

\begin{table*}[tb]
\caption{Dust properties}
\label{dusttab} 
\centering
\scriptsize
\begin{minipage}{0.85\linewidth}
\begin{tabular*}{1.0\linewidth}%
{@{\extracolsep{\fill}}|l|c|rr|cc|ccc|}\hline\hline
Object&	 dust morphology& \multicolumn{2}{c|}{IRAS flux (mJy)}& $\rm Log(M_{d,IRAS})$&	$\rm Log(M_{d,optical})$&	$\rm T_d$& $\dot{\mathcal{M}}$& $\tau_d$\\
&	&	$\rm 60\pm err$&	$\rm 100\pm err$&	$\rm (M_{\sun})$&	$\rm (M_{\sun})$& 	(K)&	$\rm (M_{\sun}yr^{-1})$&($10^6$ yr)\\
(1)& (2)& (3)& (4)& (5)& (6)& (7)& (8)& (9)\\
\hline
NGC 524&    dust rings&	780$\pm$34&	1820$\pm$114&	5.95$\pm$0.11&	4.91$\pm$0.14&	35$\pm$4& 1.06& 7.8\\
NGC 984&    maj. dust lane&	140$\pm$37&	120$\pm$106&	4.85$\pm$0.88&	4.50$\pm$0.07&	$<30$& $<0.89$& 5.5\\
NGC 1172&   nucl. dust& 	$0\pm 41$&	$0\pm72$&	$<4.35$&	4.16$\pm$0.18&	--& $<0.12$& 13.5\\
NGC 1439&   min. dust lane& 	$0\pm32$&	300$\pm$35&	$<5.80$&	5.43$\pm$0.19&	$\le20$& $<0.04$& 23.0\\
NGC 2128&   min.dust lane&	870$\pm$37&	2160$\pm$166&	6.22$\pm$0.11&	4.42$\pm$0.12&	34$\pm$3& 0.10& --\\
NGC 2534&   min.dust lane&	380$\pm$50&	800$\pm$184&	5.71$\pm$0.28&	4.10$\pm$0.18&	36$\pm$4& 0.15& --\\
NGC 2563&   min. dust lane&	        $0\pm27$&	$0\pm187$&	$<4.70$&	3.75$\pm$0.15&	--&  $<0.51$& 3.3\\
NGC 2672&   nucl. dust&	$0\pm45$&	440$\pm$45&	$<6.30$&	5.30$\pm$0.30&	$\le20$& $<0.69$& --\\
NGC 2693&   min dust lane&	210$\pm$37&	790$\pm$114&	6.09$\pm$0.20&	4.84$\pm$0.15&	29$\pm$4& 0.02& 5.8\\
NGC 2907&   multi. lanes&	310$\pm$26&	1090$\pm$194&	5.70$\pm$0.21&	5.05$\pm$0.22&	30$\pm$2& 0.47& --\\
NGC 2911&   maj. dust lane&	290$\pm$28&	560$\pm$79&	5.59$\pm$0.19&	5.54$\pm$0.14&	38$\pm$5& 0.10& 8.8\\
NGC 3078&   nucl. dust&	--&	--&	--&	3.91$\pm$0.08&	--& --& 11.0\\
NGC 3489&   dust clouds&	--&	--&	--&	3.94$\pm$0.26&	--& --& 18.5\\
NGC 3497&   multi. lanes&	278&	$<1347$& $<6.49$&	5.08$\pm$0.14&	$\le26$& --&-- \\
NGC 3585&   nucl. dust&	160$\pm$42&	$0\pm70$&	$<4.24$&	3.73$\pm$0.25&	$\le30$& 0.37& 14.7\\
NGC 3599&   min. dust lane&	--&	--&	--&	3.73$\pm$0.23&	--& --& 24.0\\
NGC 3665&   maj. dust lane&	1960$\pm$40&	6690$\pm$163&	6.60$\pm$0.07&	5.02$\pm$0.28&	30$\pm$3& 0.19& 23.1\\
NGC 3923&   nucl. dust&	$0\pm35$&	$0\pm120$&	        $<4.70$&	4.46$\pm$0.08&	--& 0.61& 10.6\\
NGC 4459&   dust ring&	1920$\pm$67&	4280$\pm$119&	        5.71$\pm$0.12&	3.66$\pm$0.14&	36$\pm$8& 0.48& 23.2\\
NGC 5363&   multi. lanes&	1700$\pm$46&	4450$\pm$45&	5.72$\pm$0.18&	5.05$\pm$0.14&	33$\pm$13& 0.01& 23.1\\
NGC 5485&   min. dust lane&	150$\pm$34&	850$\pm$88&	6.01$\pm$0.17&	4.36$\pm$0.11&	25$\pm$4& 0.16& 13.0\\
NGC 5525&   maj. lane + ring&	--&	--&	--&	5.99$\pm$0.10&	--&  --& --\\
NGC 5898&   min dust lane&	130$\pm$37&	200$\pm$64&	4.85$\pm$0.41&	4.48$\pm$0.07&	41$\pm$9& 0.25& 5.4\\
NGC 5903&   nucl. dust&	$0\pm23$&	$0\pm112$&	        $<6.31$&	4.05$\pm$0.08&	--& 0.43& 5.2\\
NGC 7432&   min. dust lane&	--&	--&	--&	3.98$\pm$0.09&	--& --& 5.8\\
NGC 7722&   multi. lanes&	820$\pm$34&	2840$\pm$159&	6.82$\pm$0.08&	6.19$\pm$0.16&	30$\pm$2& 1.72& --\\
\hline		   
\end{tabular*}
\scriptsize
\begin{list}{}{}
\item[]\textsc{Notes to Table ~\ref{dusttab}:} Column(2) lists morphology of dust: dust ring(s) - face on dust ring(s); maj.dust lane - dust lane along the optical major axis; min. dust lane - dust lane aligned along minor axis; multi. lanes - multiple dust lanes;  dust clouds - dust distribution in the form of clouds; nucl. dust - nuclear dust. Column (3) \& (4) list IRAS flux densities at $\rm 60\mu m$ and $\rm 100\mu m$ (taken from Knapp et al. 1989). Column (5) and (6) list dust masses derived from IRAS flux densities ($\rm M_{d,IRAS}$) and optical extinction values ($\rm M_{d,optical}$), respectively. Column (7) gives dust temperature ($\rm T_d$), while column (8) lists the present day mass-loss rate from the evolved, red-giant stars in the program galaxies. Column (9) lists  life time of dust grains derived using Eq.(~\ref{tau-d}).
\end{list}
\end{minipage} 
\end{table*}
\subsection{Dust mass estimation}
We have  estimated dust content of the program galaxies using (i) optical extinction studied here and (ii) IRAS fluxes at 60 $\mu$m and 100 $\mu$m taken from Knapp et al. (1989), as described in the following subsections.

\subsubsection{Using optical extinction}
To estimate dust mass using total optical extinction, we followed the method described by Goudfrooij et al. (1994c),which is outlined here. For a given grain size distribution function $n(a)$ of the spherical grains of radius $a$ and extinction efficiency $Q_{ext}(a,\lambda)$, the extinction cross-section at wavelength $\lambda$ is given by
\begin{equation}
C_{ext}(\lambda)=\int_{a_-}^{a_+}Q_{ext}(a,\lambda)\, \pi\, a^{2}\,n(a)\,da
\end{equation}
where $a_-$ and $a_+$ are the lower and upper cutoffs of the grain size distribution, respectively. Assuming $n(a)$ to be the same over the entire dusty region and using the definition of the efficiency factor, $Q_{ext}(a,\lambda)= C_{ext}(a,\lambda)/\pi a^2$ (ratio of extinction cross-section to the geometrical cross-section), the total extinction due to dust at wavelength $\lambda$ is expressed as
\begin{equation}
A_{\lambda}=1.086\,\, C_{ext}({\lambda})\times l_{d}
\end{equation}
where $l_d$ is the dust column length along the line of sight. The column length density in units of g cm$^{-2}$ for the dust is then expressed as
\begin{equation}
{\Sigma}_d= \int_{a_-}^{a_+}\frac{4}{3}\,\pi\, a^3\, \rho_d\, n(a)\,da\times l_d
\end{equation}
where $\rho_d$ gives the specific grain mass density which is taken to be $\sim$ 3 g cm$^{-3}$ for graphite and silicate grains (Draine \& Lee 1984). This is then multiplied by the total area occupied by dust to obtain the dust mass,   $\rm M_d=\Sigma_d \, \times$ Area, expressed in solar mass units.

The measured total extinction in $V$ band ($A_V$) can be used to compute dust mass using the size distribution of Mathis et al. (1977) as
\begin{displaymath}
n(a)=n_0\,\,a^{-3.5} \hspace{10mm}     (a_- \leq a \leq a_+)
\end{displaymath}
where $a_- = 0.005\,\, \mu m$ and $a_+=0.22\,\, \mu m$ for the Milky Way with $R_V=3.1$ (cf. Draine \& Lee 1984). Since the observed extinction curves in the program galaxies refer to the dust grains at the upper end of the size distribution (Goudfrooij et al. 1994c), upper limits of the grain size in the target galaxies were scaled accordingly using the relation
\begin{equation}
a_+=\frac{<a>}{a_{\rm Gal}} \times 0.22 \,\, \mu m 
\end{equation}
where $\frac{<a>}{a_{\rm Gal}}$ is the relative grain size for the program galaxies and are listed in Table ~\ref{Rvtab}.\\
Further, to estimate the extinction efficiency of the dust grains we assume spherical grains composed of silicate and graphite with nearly equal abundance (see, Mathis et al. 1977). The values of the extinction efficiency for silicate and graphite grains are taken as
\begin{displaymath}
Q_{ext,silicate} = \left\{ \begin{array}{ll} 
 0.8\, a/{a_{silicate}} & \textrm{for $a < a_{silicate}$},\\
0.8 & \textrm{for $a \geq a_{silicate}$}
\end{array} \right.
\end{displaymath}
\begin{displaymath}
Q_{ext,graphite}=\left\{ \begin{array}{ll} 
2.0\, a/{a_{graphite}} & \textrm{for $a < a_{graphite}$},\\ 
2.0 & \textrm{for $a \geq a_{graphite}$}

\end{array} \right.
\end{displaymath}
with $a_{silicate}=0.1\, \mu$m, and $a_{graphite}=0.05\, \mu$m. 
Using these parameters and the dust column density, the total dust mass ($\rm M_{d,optical}$) contained in the program galaxies was estimated. In determining the mean visual extinction, we included all those regions with $\tau_V \ge0.02$. The computed dust masses for the sample galaxies using total optical extinction ($\rm M_{d, optical}$) are given in column 6 of Table~\ref{dusttab}. 

\subsubsection{Using IRAS densities}
Using IRAS flux densities measured at 60 and 100 $\mu$m, we first calculate the dust grain temperature in the program galaxies using relation, $\rm T_d \, = 49\left(\frac{S_{60}}{S_{100}}\right) ^{0.4}$ (Young et al. 1989). The  dust content ($\rm M_{d,IRAS}$) in the sample galaxies was then computed using the relation (Hildebrand 1983):
\begin{equation}
M_d = \frac{4}{3}\, a \,\rho_d\, D^2 \frac{F_\nu}{Q_\nu B_\nu (T_d)}
\end{equation}
where $a, \, \rho_d$ and  $D$ are the grain radius, specific grain mass density and distance of the galaxy, respectively. $F_\nu, \, Q_\nu \,$\& $B_\nu (T_d)$ are the observed flux density, grain emissivity and the Planck function of temperature $\rm T_d$ at frequency $\nu$, respectively. We adopted  $\rho_d = 3$ g\,$cm^{-3}$  and $\frac{4a\,\rho_d}{3Q_\nu}$ = 0.04 g\,$cm^{-2}$ for 0.1 $\mu$m grains at 100 $\mu$m (Hildebrand 1983).  The derived dust masses from IRAS flux densities are listed in column 5 of the Table 5. Due to the fact that IRAS was insensitive to the dust emitting at wavelengths longer than 100 $\mu$m (i.e., dust cooler than about 20\,K), these estimates of dust masses using IRAS flux densities represent the lower limits (Tsai \& Mathews 1996). 

\section{Discussion}
\begin{description}
\item[] Extinction curves derived for the sample galaxies reveal that the $R_V$ value lies in the range between 2.03 to 3.46 for ellipticals, and 2.25 to 3.46 for lenticulars, with an average of 3.01, compared to 3.1 in the case of the Milky Way. For some of the sample galaxies the $R_V$ value is close to the classical Galactic value, while for some of them the $R_V$ value is significantly less than 3.1. Interestingly, all galaxies  (e.g., NGC 2534, 2907, 2911, 5363, 5485 and 7722) showing smaller $R_V$ values exhibit well-settled dust lane or lanes with an average $R_V$  equal to 2.80, implying that the ``larger'' grains responsible for the optical extinction are roughly 25\% ``smaller'' in size compared to the grains responsible for the extinction curve in the Milky Way. For a sample of 10 elliptical galaxies, Goudfrooij et al. (1994c) arrived at  $R_V$ in the range of 2.1 to 3.3, with characteristic dust grains up to 30\% smaller in size compared to the canonical grains in the Milky Way. Thus, our results are in good agreement with those of Goudfrooij et al. (1994c). For a few of the remaining galaxies (e.g., NGC 524, 2672, 2693, 3489, 3585) extinction curves are found to lie slightly above the Galactic extinction curve, implying marginally larger values of $R_V$, and these are found to have ring or arc shaped dust morphologies. Thus, our results seems to indicate that the galaxies having $R_V$ values smaller than the canonical value (and hence smaller grains) exhibit smooth, regularly distributed dust lanes, whereas the galaxies with  larger $R_V$ values exhibit irregular dust morphologies (Goudfrooij et al. 1994c).
\item[] Dust masses derived from optical extinction are found to lie in the range $\rm\sim10^4$ to $\rm 10^6 \, M_{\sun}$ and are in good agreement  with the earlier estimates for the early-type galaxies (Brosch et al. 1990; Goudfrooij et al. 1994c; Sahu et al. 1998; Ferrari et al. 1999; Dewangan et al. 1999; Tran et al. 2001; Patil et al. 2002). The dust masses derived using the optical methods depend on the assumption of a foreground screen in front of the  background light. Therefore, this method is not sensitive to any dust embedded or intermixed with the stars within the galaxy and hence the optical method always provides a lower limit of the true dust content of the host galaxy. Comparison of the dust masses derived from IRAS flux densities and optical extinctions reveals that $\rm M_{d,IRAS}$ is larger than $\rm M_{d,optical}$ by a factor of 7.8 for the sample of  galaxies studied here. Goudfrooij \& de Jong (1995) for a sample of elliptical galaxies alone reported this ratio to be equal to 8.4. In the case of S0 galaxies, as pointed out by Sahu et al. (1998) and Dewangan et al. (1999), this mass discrepancy is less significant compared to that for ellipticals. Using ISO data Temi et al. (2004) estimated a dust content of NGC 1172 and NGC 5363 of $9.9\times 10^5\,M_{\sun} \, \& \, 2.00\times 10^6\,M_{\sun}$, respectively, these estimates are roughly an order of magnitude larger than our estimates from IRAS fluxes. This in turn implies that although ISO data acts as a more reliable tracer of the true dust content in the extragalactic environment, it renders the dust mass discrepancy even more significant than using IRAS flux densities.

\item[] There are several processes that may alter the effective dust grain size in the external galaxies, like destruction of grains due to  sputtering in supernova blast waves, grain-grain collisions, sputtering by thermal ions (warm and hot) etc., as has been discussed by Goudfrooij (1999). But, which one of these would work as the dominant process for destruction of grains leading to smaller grain size depends on the conditions of specific galaxies. In particular, for galaxies immersed in hot X-ray emitting gas, sputtering by hot ions is expected to be the dominant process for destruction and alteration of the grain size. In order to investigate this point further we collected X-ray flux for the sample galaxies from the literature, and examined the relationship between the dust mass and X-ray luminosity, both normalized to the blue luminosity. The result shown in  Fig.~\ref{Xraydust}(a) reveals a marginal anti-correlation suggesting that high ($L_X/L_B$) galaxies harbour less dust, in agreement with the predictions of Goudfrooij et al. (1994c).

\begin{figure*}[!t]
\centering
\begin{minipage}[r]{0.5\textwidth}
\centering\includegraphics[width=3.0in]{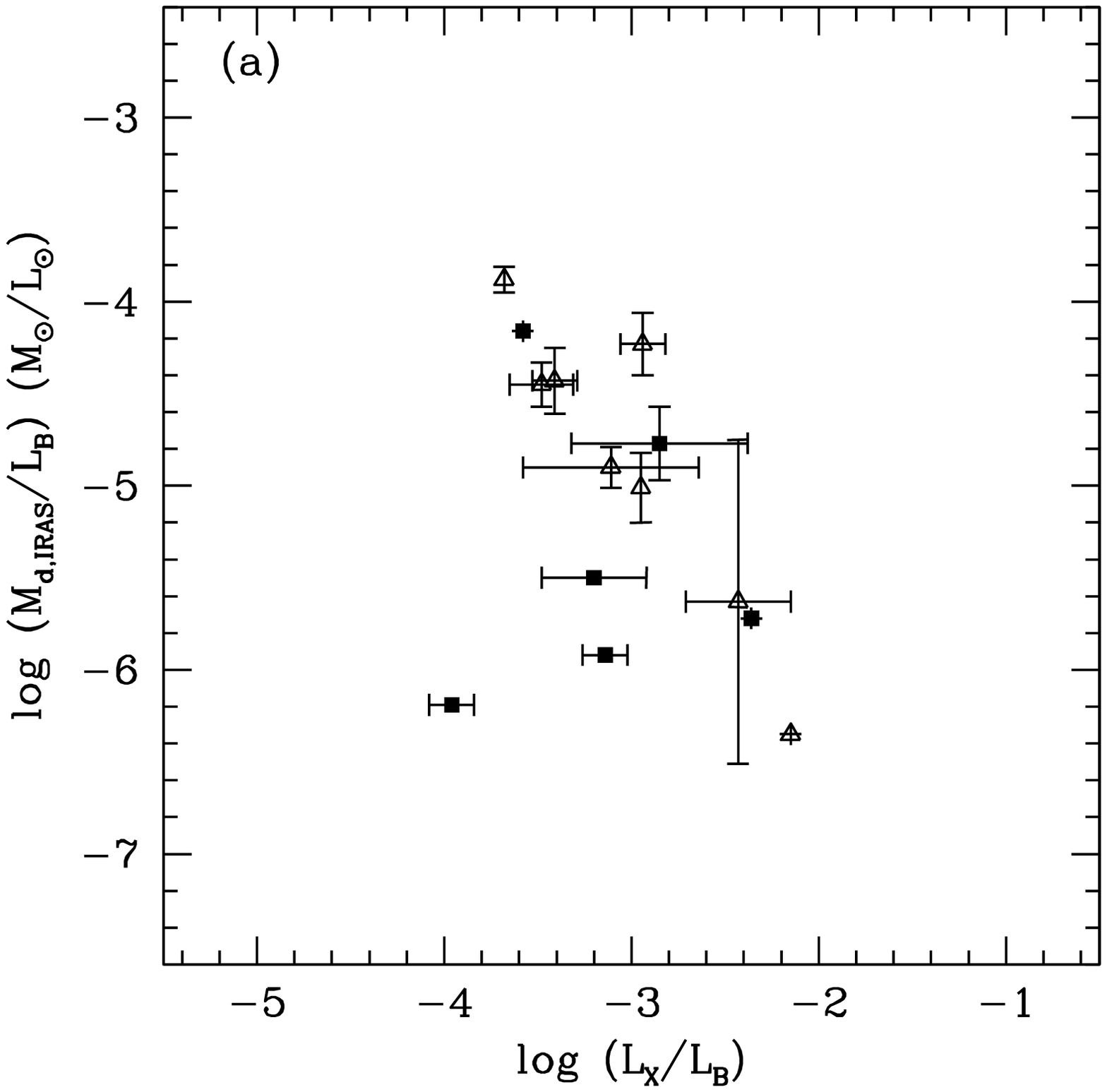}
\end{minipage}%
\begin{minipage}[l]{0.5\textwidth}
\centering\includegraphics[width=3.0in]{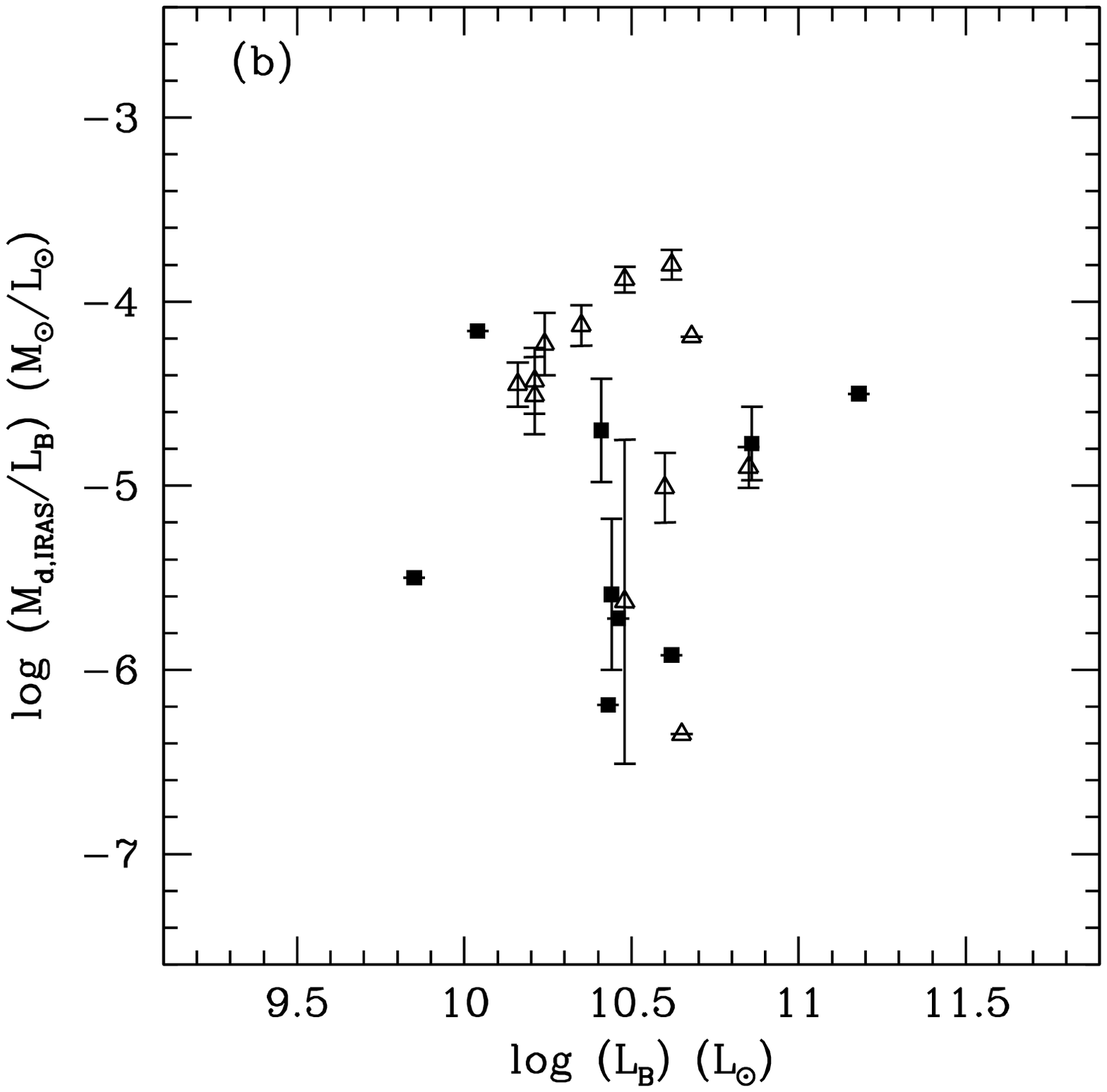}
\end{minipage}
\caption{The ratio of dust mass to blue luminosity $\left(\frac{M_{d,IRAS}}{L_B}\right)$ as a function of (a) ratio of the X-ray-to-blue luminosity, and (b) blue luminosity (filled squares are for E, while open triangles for S0).} 
\label{Xraydust}
\end{figure*}

\item[] The issue of the origin of dust in early-type galaxies is rather controversial. Suggested sources by which galaxies may acquire dust include (i) internal: in the form of mass-loss from evolved stars distributed throughout the galaxy more or less uniformly, and (ii) external: in which a gas-rich companion donates gas and dust either via direct merger, accretion or tidal capture. Dwek \& Scalo (1980) have demonstrated that evolved red giant stars are the dominant source of dust injection into the ISM of the early-type galaxies. To check the significance of internal sources, using the present day mass-loss rate we estimated the total amount of dust accumulated over the lifetime of a galaxy under the assumption that the properties of the  evolved, red-giant stars in the Galactic bulge closely resemble those in elliptical galaxies.  For most of the K \& early M giant stars with effective surface temperatures about 4500\,K and with no circumstellar material, one finds  $S_{12\,\mu m}/S_{2.2\,\mu m}$ = 0.08 (Knapp et al. 1992), while for the stars with circumstellar matter (as in elliptical galaxies) this ratio is 0.13, about $38\%$ larger than expected from stars in the Galactic bulge (Knapp et al. 1992). This excess flux measured at 12 $\mu$m in elliptical galaxies is attributed to the emission from the circumstellar shells produced by mass-loss from the evolved stars, and thus provides a reasonable  estimate for the total mass-loss. Following Knapp et al. (1992), we adopt the following relation for the mass-loss rate: 

\begin{equation}
\dot{\mathcal{M}} = 6\times 10^{-6} \, \bigg(\frac{D}{Mpc}\bigg)^2 \Bigg(\frac{S_{12}- 0.042\, S_{100}}{mJy}\Bigg) \, M_{\sun}\,  yr^{-1}.
\end{equation} 

The term 0.042 S$_{100}$ gives the correction to the excess flux at 12 $\mu$m emission due to contamination from the interstellar dust heated to an equilibrium temperature of about 20\,K by the ambient ultraviolet interstellar radiation field. The inferred values of the mass-loss rates in the sample galaxies are listed in column 8 of Table~\ref{dusttab}. 

\item[] Dust injected into the ISM by the evolved stars through mass-loss is also destroyed simultaneously by a variety of processes. In the absence of hot, X-ray emitting gas, the life time of the 0.1$\,\mu$m refractory grains against  sputtering through low velocity shock waves or grain-grain collision is $\sim10^9$ yr, while that in the presence of hot coronal gas ($\rm T_e \sim 10^7$ K) against  collision with the hot protons and $\alpha$-particles is of the order of $\sim 10^6 - 10^7$ yr (cf. Draine \& Salpeter 1979, Goudfrooij \& de Jong 1995). A majority of the galaxies from the present sample are detected as X-ray emitters, and therefore, we estimate the lifetime of the dust grains (column 9 of Table~\ref{dusttab}) of radius $a$ against such collision using the relation 
\begin{equation}
\label{tau-d}
\tau_d = a\left|\frac{da}{dt}\right|^{-1}\simeq 2\times10^4 \left(\frac{cm^{-3}}{n_H}\right) \left(\frac{a}{0.01\,\mu m}\right)\,\,\,\,  yr
\end{equation}
(Draine \& Salpeter 1979), where $n_H$ is the proton density in the plasma and $a$ is the grain size. Here $n_H= 0.83\, n_e$, $n_e$ being the electron density estimated using 
\[
n_e(r)=n_e(0)\left[1+\left(\frac{r}{a_x}\right)^2\right]^{-3/4}
\]
 and 
\[n_e(0) = 0.061\left(\frac{L_X}{10^{41} \rm{ergs\, s^{-1}}}\right)^{1/2}\left(\frac{a_x}{1\, \rm{kpc}}\right)^{-3/2} cm^{-3}
\]
 where $n_e(0)$ is the central number density over the X-ray core radius $a_X\,\big(= 1.0\left(\frac{L_B}{10^{11}\, L_{\odot}} \right)^{0.8} \rm{kpc}\big)$ (see, e.g., Canizares et al. 1987, for details). X-ray luminosities reported by Bettoni et al. (2003) were used to derive these parameters. For the grains of radius 0.1 $\mu$m this time scale is found to lie in the range between $2.4 \times 10^7\,\rm{yr}$ and $3.3 \times 10^6\,\rm{yr}$ which correspond to  NGC 3599 and NGC 2563 having the lowest and highest X-ray luminosities, respectively, in the present sample. 
\begin{figure}[!b]
\centering
\includegraphics[width=3.2in, height=2.6in, trim=0 60 20 40, clip]{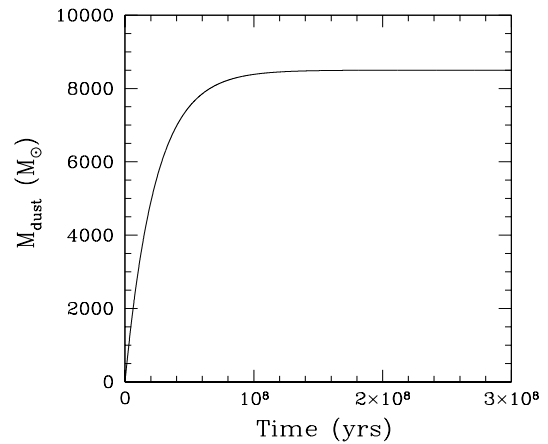}
\caption{Total build up of dust mass within NGC 5363 over its life time}
\label{dustbuildup}
\end{figure}

For illustration, we examine the accumulation of dust for the case of NGC 5363 (log $L_X$= 40.38 ergs s$^{-1},\, n_H=2.2\times 10^{-3} cm^{-3}, \tau_d=2.3\times 10^7$ yr and $\dot{\mathcal{M}}=10^{-2}\,\rm{M_{\odot}\, yr^{-1}}$ and gas-to-dust ratio of about 100), taking into account the two competing processes, namely mass-loss in the form of dust from the evolved stars and destruction of dust. The rate of accumulation of dust is
 \[\frac{\partial M_d(t)}{\partial t}=\frac{\partial M_{d,s}}{\partial t} - M_d(t)\,\,\tau_d^{-1}\,\,, \]
where $\frac{\partial M_{d,s}}{\partial t}$ is the  rate of mass loss  in the form of dust from red giant stars and $\tau_d^{-1}$ is the destruction rate. Integrating this dust accumulation rate over time, the total build up of dust mass in NGC 5363 over its life time is shown in Fig.~\ref{dustbuildup} and is found to be equal to 8.5 $\times 10^3\, \rm{M_{\odot}}$, which is about a factor of 100 times smaller than the observed dust mass. For other galaxies, where X-ray flux was available in the literature, comparison of the build up mass and the observed mass shows a similar discrepancy except for NGC 3585 and NGC 3923, for which the two estimates are roughly equal. This in turn implies that internal sources are not sufficient to account for the observed dust and therefore an external source plays a dominant role in the supply of dust into the ISM of early-type galaxies.

\item[] There are several other indicators that support the external origin of dust in the early-type galaxies. If it is of internal origin, it should exhibit the same global rotational dynamics as that of the stellar system (Kley \& Mathews 1995). However, kinematical studies of dust lane galaxies have shown that the angular momentum vectors of the interstellar gas are often orthogonal to those of the stars, implying that the ISM must have a different origin (Caon et al. 2000). Another good argument in favour of an external, merger-related origin of dust in early-type galaxies is the lack of significant correlation between the $L_{FIR}$ or dust mass ($\rm M_{d,IRAS}$) and optical luminosity ($L_B$), Fig.~\ref{Xraydust}(b). This suggests that at least some of our sample galaxies have acquired dust externally in a merger-like event. Similar conclusions were also arrived at by Forbes (1991), Goudfrooij \& de Jong (1995),  Temi et al. (2004). 

\begin{figure}[!htb]
\includegraphics[height=2.42in, width=3.5in, trim=0 10 30 180, clip]{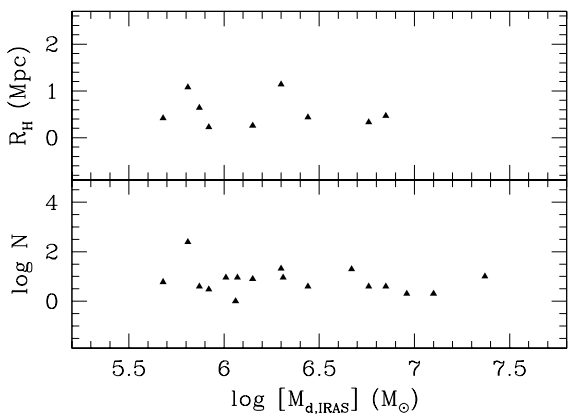}
\caption{Environmental dependence of dust mass; upper and  lower panels represent dependence on group members (N) and group harmonic radius ($R_H$), respectively}
\label{group}
\end{figure}

\item[] To explore the possibility that dust could originate through a direct merger or accretion or tidal capture of gas from their encounter with neighbouring galaxies, we searched the environment of the program galaxies and found  that the majority of the galaxies are either a part of a group\,/\,cluster, or form interacting pairs with their neighbours. Relevant details on the environment of the galaxies taken from Huchra \& Geller (1982), Geller \& Huchra (1983) and Garcia (1993)  are given in the column 8 of Table~\ref{obslog}. Following Tran et al. (2001), we also examined the relationship between the dust content and environment of the program galaxies using  parameters like group membership ($N$) and group harmonic radius ($R_H$); $R_H$ provides a rough measure of the size of the group (see, Tran et al. 2001 for details). The group membership and group harmonic radius is plotted  as  a function of dust mass in Fig.~\ref{group}. The scatter plots between $M_d$ vs. $N$ \& $R_H$ indicate the absence of any significant correlation between dust mass and the two quantities. As group membership in these catalogs mostly relies on the radial velocity measurement of high-luminosity galaxies, small (dwarf) galaxies are often missing from these catalogs, and hence log N  may not be representative of the real situation. If the donors of the dust are supposed to be galaxies similar to those present in the group catalogs mentioned above, this would make the possibility of dust originating from neighbouring companions through merger or tidal capture more complicated. However, one should interpret these results with caution, as a given dusty galaxy might have acquired dust sufficiently long ago that no signature of merger or interaction is visible at the present epoch (Tran et al. 2001).
\end{description}

\section{Conclusions}

We have reported properties of dust in a sample of 26 early-type galaxies based on their multicolour imaging observations. Our results are summarized here.

Extinction curves derived for galaxies studied here run parallel to the canonical curve of the Milky Way, implying that properties of dust in the extragalactic environment are similar to those of the canonical grains in the Milky Way. The $R_V$ value, which characterizes the extinction curve in the optical region, is found to vary in the range 2.03 - 3.46 with an average of 3.02, and is not very much different from the canonical value of 3.1 for our Galaxy. However, several galaxies studied here exhibit smaller $R_V$ values suggesting that the ``large'' grains responsible for the optical extinction are significantly smaller in size than the canonical grains in the Milky Way. Our results indicate that $R_V$ values largely depend on the morphology of dust in the host galaxies, in the sense that galaxies with well-defined dust lanes exhibit much smaller grains compared to those with irregular dust morphologies.

The dust content derived by the optical extinction method for  the program galaxies is found to lie in the range $\rm 10^4$ to $\rm 10^6\, M_{\sun}$ and is, in general, less than the  dust mass derived from IRAS flux densities. This discrepancy is more significant in the case of ellipticals than lenticulars, in good agreement with the results obtained by other researchers for early-type galaxies.

Galaxies with well settled dust lanes contain high dust masses and are weak X-ray sources, in agreement with  the prediction of Goudfrooij (1994c). Further, galaxies with well-settled dust lanes have smaller values of $R_V$, and are found to lie in denser environments.

For origin of dust in early-type galaxies, internal mass-loss from evolved stars alone cannot account for the observed dust in these galaxies and therefore points towards an external origin for the dust in these galaxies. 
\begin{acknowledgements}
The authors thank Drs. Ram Sagar, Vijay Mohan, Mahendra Singh and the staff of ARIES, VBO, IAO for their help during the observing runs. We are grateful to the anonymous referee whose insightful comments helped us to improve the paper. We also acknowledge the enlightening discussion with Dr. Paul Goudfrooij during this work. MKP \& SKP are grateful to IUCAA for hospitality and for use of their computational and library facilities. MKP thanks Padmakar, Sudhanshu, Ravi Kumar and Laxmikant Chaware for their valuable help and discussion during this work. We acknowledge the use of  NASA/IPAC Extragalactic Database (NED).
\end{acknowledgements}


\begin{thebibliography}{}
\bibitem[]{} Bettoni, D., Galletta, G., \& Burillo, S.G. 2003, A\&A, 405, 5
\bibitem[]{} Brosch, N. 1988, in Dust in the universe, Cambridge Univ. Press, 501  
\bibitem[]{} Brosch, N., Almoznino, E. 1997, ApLC, 35, 371
\bibitem[]{} Brosch, N., Almoznino, E, Greenberg J.M., Grosbel P.G. 1990, A\&A, 233, 341  
\bibitem[]{} Brosch, N. \& Loinger F. 1991, A\&A, 249, 327
\bibitem[]{} Canizares, C.R., Fabbiano, G. \& Trinchieri, G. 1987, 312, 503
\bibitem[]{} Caon, N., Macchetto, D., Pastoriza, M. 2000, ApJSS, 127, 39
\bibitem[]{} Clayton, G.C., Wolff, M.J., Sofia, U.J., Gordon, K,D.,\& Misselt, K.A. 2003, ApJ, 588, 871
\bibitem[]{} de Vaucouleurs, G., de Vaucouleurs, A., Corwin, H.G., Buta, R., Paturel, G. \& Fouque, P. 1991, \emph{Third Reference Catalog of bright galaxies}, New York:Springer (RC3)
\bibitem[]{} Dewangan, G.C., Singh, K.P., Bhat, P.N. 1999, AJ, 118, 785
\bibitem[]{} Draine, B. T. \& Lee, H. M. 1984, ApJ, 285, 89
\bibitem[]{} Draine, B. T. \& Salpeter, E. 1979, ApJ, 231, 77
\bibitem[]{} Dwek, E. \& Scalo, J.M. 1980, ApJ, 239, 193
\bibitem[]{} Ebneter, K., \& Balick, B. 1985, AJ, 90, 183
\bibitem[]{} Falco, E. E., Impey, C. D., Kochanek, C. S., Lehár, J., McLeod, B. A., Rix, H.-W., Keeton, C. R., Muñoz, J. A., Peng, C. Y. 1999, ApJ, 523, 617
\bibitem[]{} Ferrari, F., Pastoriza, M.G., Macchetto, F., \& Caon N. 1999, A\&ASS, 136,269
\bibitem[]{} Forbes, D.A. 1991, MNRAS 249, 779
\bibitem[]{} Garcia, A.M. 1993, A\&AS, 100, 47
\bibitem[]{} Geller, M.J. \& Huchra, J.P. 1983, ApJS, 52, 61
\bibitem[]{} Goudfrooij, P. 1999, Proceedings of IAU Colloquium 174, ASP Conf. Series, eds. M. Valtonen \& C. Flynn
\bibitem[]{} Goudfrooij, P. \& de Jong ,T. 1995, A\&A 298, 784
\bibitem[]{} Goudfrooij, P., \& Trinchieri, G. 1998, A\&A 330, 123
\bibitem[]{} Goudfrooij, P., Hansen, L., Joregensen, H.E., Norgaard-Nielsen, H.U. 1994b, A\&A SS 105, 341
\bibitem[]{} Goudfrooij, P., de Jong ,T., Hansen, L.,  Norgaard-Nielsen, H.U. 1994c, MNRAS 271, 833
\bibitem[]{} Hildebrand, R.H. 1983, QJRAS, 24, 267
\bibitem[]{} Huchra, J.P. \& Geller, M.J. 1982, ApJ, 257, 423
\bibitem[]{} Jedrzejewski, R. I. 1987, MNRAS 226, 747
\bibitem[]{} Keel, W. C. \& White, R.E. 2001, AJ, 121, 1442
\bibitem[]{} Kley, W., \& Mathews, W.G. 1995, ApJ, 438, 100
\bibitem[]{} Knapp, G. R., Guhathakurta, P., Kim, D.-W., Jura, M. 1989, ApJS 70
, 329
\bibitem[]{} Knapp, G. R., Gunn, J.E. \& Wynn-Williams, C.G. 1992, ApJ, 399, 76
\bibitem[]{} Landolt, A.U. 1992, AJ, 104, 340
\bibitem[]{} Leeuw, L.L., Sanson, A.E., Robson, E.I., Haas, M., Kuno, N. 2004, ApJ, 612, 837
\bibitem[]{} Maiolino, R., Schneider, R., Oliva, E., Bianchi, S., et al. 2004, Nature, 431, 533
\bibitem[]{} Massa, D., Savage, B., Fitzpatrick, E.L. 1983, ApJ, 266, 662
\bibitem[]{} Mathis, J.S. 1990, ARA\&A, 28, 37
\bibitem[]{} Mathis, J.S., Rumpl, W., Nordsieck, K.H. 1977, ApJ, 217, 425 
\bibitem[]{} Motta, V., Mediavilla, E., Mutoz, J. A., Falco, E., Kochanek, C. S., Arribas, S., García-Lorenzo, B., Oscoz, A., Serra-Ricart, M. 2002, ApJ, 574, 719
\bibitem[]{} Patil, M.K., Sahu, D.K., Pandey, S.K., Kembhavi, A.K., Joshi, U.C., Baliyan, K.S. \& Singh,M. 2002, \emph{Bull. Astr. Soc. India}, 30, 759
\bibitem[]{} Rieke, G.H. \& Lebofsky, M.J. 1985, ApJ, 288, 618
\bibitem[]{} Sahu, D.K., Pandey, S.K., Kembhavi, A.K. 1998, A\&A, 333, 803
\bibitem[]{} Savage, B.D. \& Mathis, J.S. 1979, ARA\&A, 17, 73.
\bibitem[]{} Temi, P., Brighenti, F., Mathews, W.G., \& Bregman, J.D. 2004, ApJS, 151, 237
\bibitem[]{} Tran, H.D., Tsvetanov, Z., Ford, H.C., \& Davis, J. 2001, AJ, 121, 2928
\bibitem[]{} Trinchieri, G., Noris, L., di Serego Alighieri, S. 1997, A\&A, 326, 565
\bibitem[]{} Tsai, J.C. \& Mathews, W.G. 1996, ApJ, 468, 571
\bibitem[]{} Valencic, L., Clayton, G.C. \& Gordon, K.D. 2004, ApJ, 616, 912
\bibitem[]{} V\'eron-Cetty, M.P., V\'eron, P. 1988, A\&A, 204, 28
\bibitem[]{} van Dokkum, P.G.,  Franx, M. 1995, AJ, 286, 415
\bibitem[]{} Xilouris, E. M., Madden, S. C., Galliano, F., Vigroux, L., \& Sauvage, M. 2004, A\&A, 416, 41
\bibitem[]{} Young, J.S., Xie, S., Kenney, J.D.P., Rice, W.L. 1989, ApJS, 70, 699
\end{thebibliography}
\end{document}